\documentclass{amia}
\usepackage{graphicx}
\usepackage[labelfont=bf]{caption}
\usepackage{amsmath,amssymb} 
\usepackage{epsfig}
\usepackage{csquotes}
\usepackage{subcaption}
\usepackage[superscript,nomove]{cite}
\usepackage[super,sort&compress]{natbib}
\setcitestyle{citesep={,}}
\usepackage{xcolor}
\usepackage{bm}
\usepackage[pagebackref=true,breaklinks=true,letterpaper=true,colorlinks,bookmarks=false]{hyperref}

\DeclareMathOperator{\E}{\mathbb{E}}

\newcommand{\update}[1]{\textcolor{black}{#1}}

\begin{document}

\title{Towards Better Opioid Antagonists Using Deep Reinforcement Learning}

 \author{
 Jianyuan Deng, M.Phil.$^{*1}$, Zhibo Yang, M.Phil.$^{*2}$, Yao Li, B.Sc.$^{3}$, \\Dimitris Samaras, Ph.D.$^{2}$, Fusheng Wang, Ph.D.$^{1,2}$}
\let\thefootnote\relax\footnotetext{*Both authors contributed equally}

 \institutes{
     $^1$Department of Biomedical Informatics, Stony Brook University\\
     $^2$Department of Computer Science, Stony Brook University\\
     $^3$Department of Chemistry, Carnegie Mellon University
 }

\maketitle

\noindent{\bf Abstract}
\textit{Naloxone, an opioid antagonist, has been widely used to save lives from opioid overdose, a leading cause for death in the opioid epidemic. However, naloxone has short brain retention ability, which limits its therapeutic efficacy. Developing better opioid antagonists is critical in combating the opioid epidemic.
\update{Instead of exhaustively searching in a huge chemical space for better opioid antagonists, we adopt reinforcement learning which allows efficient gradient-based search towards molecules with desired physicochemical and/or biological properties.} Specifically, we implement a deep reinforcement learning framework to discover potential lead compounds as better opioid antagonists with enhanced brain retention ability. 
A customized multi-objective reward function is designed to bias the generation towards molecules with both sufficient opioid antagonistic effect and enhanced brain retention ability. Thorough evaluation demonstrates that with this framework, we are able to identify valid, novel and feasible molecules with multiple desired properties, which has high potential in drug discovery.
}

\section{Introduction}
\begin{figure}[h!]
\centering
\includegraphics[scale=0.4]{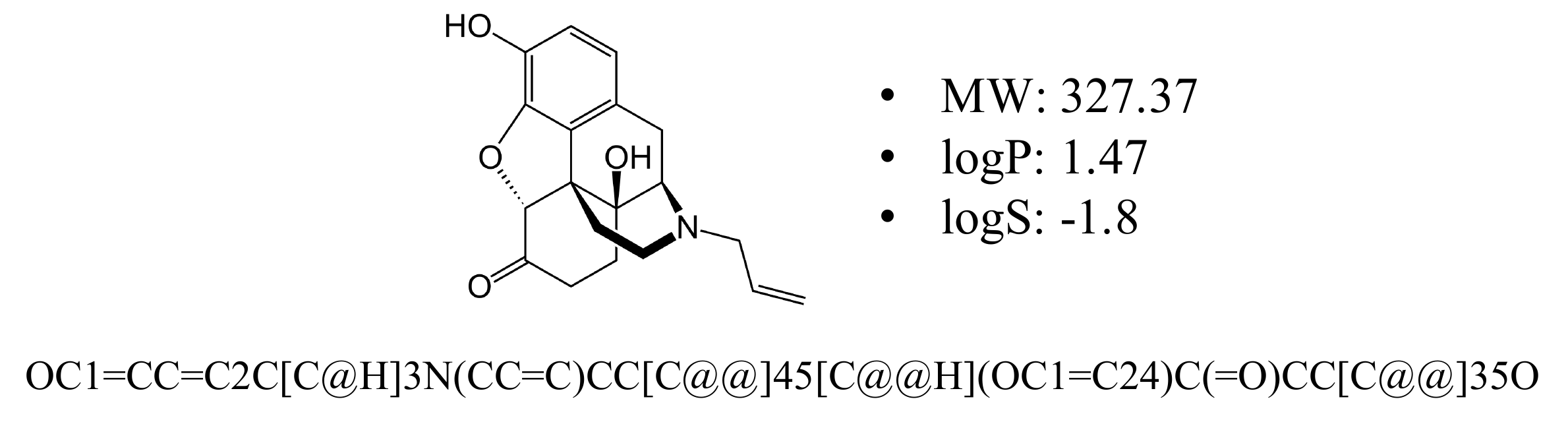}
\caption{Structure and Physicochemical Properties of Naloxone. \cite{wishart2018drugbank}}
\label{Naloxone}
\end{figure}

Over the last 20 years, there is a dramatic rise in the use and misuse of opioids in the United States, including misuse of prescription opioids, resurgence in heroin use and increase in abuse of illicit synthetic opioids such as fentanyl, which led to the current opioid epidemic and caused a rising number of overdose deaths\cite{skolnick2018opioid}. 
According to the Centers for Disease Control and Prevention, the rate of opioid overdose deaths keeps rising from 1999 to 2018, posing a major threat to  public health\cite{hedegaard2020drug}. 
Opioid overdose happens when an excessive amount of opioid agonists work on the \update{$\mu$-opioid receptor (MOR)} in the brain, resulting in respiratory depression and eventually death\cite{schiller2019opioid}. 
To reverse opioid overdoses, naloxone as shown Figure \ref{Naloxone}, an antagonist to the MOR, is used as the most common antidote, usually in the nasal formulation so as to efficiently bypass the blood brain barrier (BBB) and exert an immediate effect\cite{skolnick2018opioid}. 

However, naloxone can be distributed away from the brain rapidly, leading to a brief period of pharmacodynamic action, which is possibly caused by its limited BBB permeability\cite{clarke2005naloxone}. Considering that opioid agonists have a longer half-life, there is a risk of inadequate response or re-narcotization after a single dose of naloxone, especially in patients who have taken large doses or long-acting opioid formulations\cite{rzasa2018naloxone}. Administration of repeated doses of naloxone may be necessary if respiratory depression recurs\cite{wermeling2013response}. \update{At the same time, the price of naloxone nasal spray is high \cite{gupta2016rising}.
Other attempts to lengthen the time for reversing opioid overdose, such as combining naloxone with other opioid antagonists, have failed \cite{krieter2019pharmacokinetic}. }
For all these reasons, there is a demand for more effective opioid antagonists with enhanced brain retention ability, which corresponds to high BBB permeability. 

Nevertheless, developing new drugs costs 2.6 billion dollars on average, and can take more than 10 years \cite{chan2019advancing}. 
Drug discovery for lead compounds, i.e., promising drug candidates, requires iterative organic synthesis and screening assays, with a failure rate higher than 90\% \cite{hughes2011principles}. Recently, the increase in the amount of chemical and biomedical data has encouraged the use of `data-hungry' machine learning algorithms such as deep learning to generate and optimize molecules, which significantly accelerates the drug discovery process by reducing resources spent on wet-lab synthesis and characterization of bad lead compounds\cite{chen2018rise, elton2019deep}.
By representing molecules as simplified molecular-input line-entry system (SMILES) strings, the generation of potential drug molecules can be treated as a sequence generation problem. With reinforcement learning (RL)\cite{sutton2018reinforcement}, the generation can be biased towards molecules with certain desired properties. For example, Popova et al \update{used RL to train a molecule generator} to generate novel compound libraries with a desired physicochemical or biological property \cite{popova2018deep}. 

For naloxone, it targets the central nervous system (CNS). Several physicochemical factors underlie permeation through the BBB for CNS drugs\cite{mikitsh2014pathways}. For instance, CNS active drugs tend to have smaller molecular weight (MW). Molecules with MW less than 500 can undergo significant free diffusion and when MW increases from 200 to 450, BBB permeability \update{decreases} 100-fold. Besides, CNS drugs must have sufficient lipophilicity (measured by the  partition coefficient between octanol and water, logP) to cross the hydrophobic phospholipid bilayer of cell membranes. One example of increased BBB permeability with higher logP is that heroin (logP=2.3) exhibits much higher brain uptake than morphine (logP=0.99). Besides, solubility (measured by logS) is also an important property because successful nasal products like naloxone nasal spray usually require the active ingredient to  be highly soluble\cite{wermeling2013response}. 

The driving question, in this study, is whether there can be molecules with both sufficient opioid antagonistic effect (i.e., a higher negative logarithm of the experimental half maximal inhibitory concentration, pIC50) and enhanced brain retention ability (i.e., a smaller MW and a higher logP) while maintaining high solubility (i.e., a higher logS). \update{Given that the number of drug-like molecules is estimated to be between $10^{30}$ and $10^{60}$, routine virtual screening on existing compound libraries can not guarantee finding molecules with multiple desired properties and exhaustive searching in the huge chemical space can be prohibitively expensive\cite{popova2018deep}.} Therefore, a multi-objective deep reinforcement learning (DRL) framework is used for the discovery of better opioid antagonists.

\section{Methods}
\subsection{A Deep Reinforcement Learning Framework}
Our framework consists of three major components: 1) a generative model based on an RNN model that can generate SMILES strings; 2) a predictive model that predicts the properties of interest for a given SMILES string; and 3) an RL engine which biases the generative model towards generating SMILE strings with desired properties, the values of which are predicted by the predictive model.

\begin{figure}[b]
\centering
\includegraphics[scale=0.4]{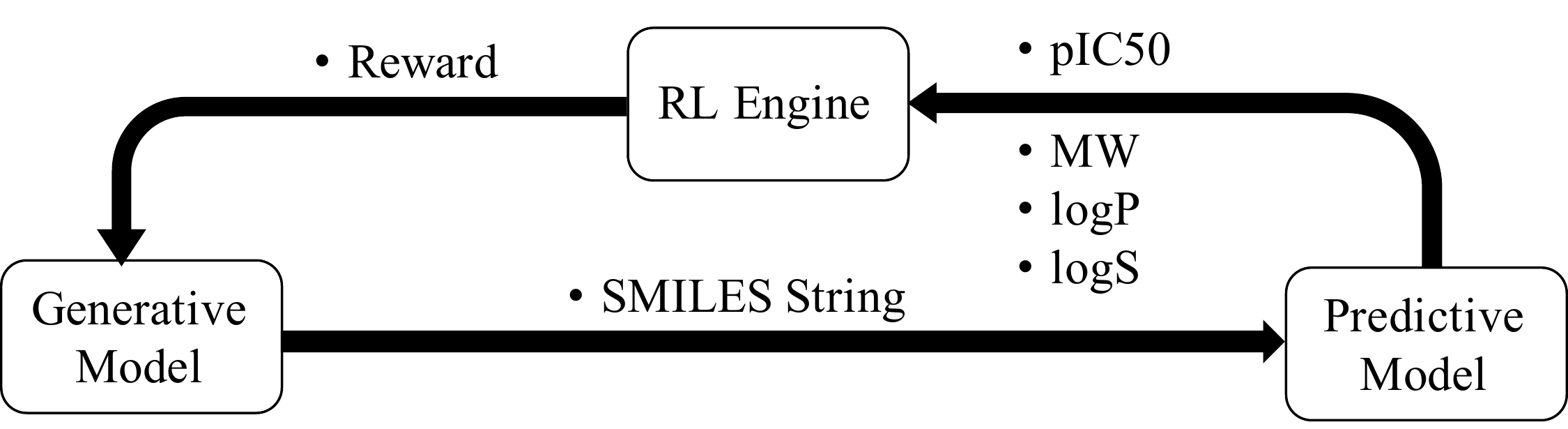}
\caption{Overview of the Deep Reinforcement Learning Framework. First, the generative model samples SMILES strings whose properties are predicted by the predictive model; the RL engine then combines all properties of each sampled SMILES into a reward as feedback to train the generative model to generate SMILES with desired properties.}
\label{fig:pipeline}
\end{figure}

Inspired by Popova et al\cite{popova2018deep}, we first train the generative model on a large corpus ($\sim$1.9M) of real-world compound SMILES strings to learn the syntax of SMILES so that the generative model is able to generate valid SMILES strings. The learned weights provide a good initialization for the generative model during the RL stage. Second, we train the predictive model which contains a predictive sub-model for every property of interest. In this paper, we built three sub-models for pIC50, logP and logS respectively. There is no sub-model for MW since it can be directly calculated. With the learned predictive model and well-initialized generative model, we use an RL algorithm, REINFORCE \cite{williams1992simple}, to further train the generative model in an end-to-end fashion such that the generated SMILES strings can have the desired properties. Figure \ref{fig:pipeline} depicts an overview of the DRL framework.

\textbf{Reinforcement Learning. }
Reinforcement learning refers to the problem of learning an optimal decision-making policy to acquire the maximal amount of rewards in a sequential decision scenario \cite{sutton2018reinforcement}. Given the previously generated SMILES characters $\bm{s}_t=\{c_i\}_{i=0}^t$ where $c_0$ is the start token, the stochastic policy (i.e., the generative model) $\pi_\theta(a|\bm{s}_t)$ samples the next SMILES character $a$ as its action. Then, a reward $r_{t+1}$ is provided by the reward function $R(\bm{s}_t, a)$ and the state $\bm{s}_t$ will be updated to $\bm{s}_{t+1}=\{c_0, \cdots, c_t, a\}$. This generation process repeats until the generative model samples a termination token or reaches a predefined maximum length of a SMILES string. RL seeks an optimal policy $\pi^*_\theta$ to maximize the expected cumulative future rewards (i.e., return)
\begin{equation}
\pi^*_\theta = \arg\max_{\pi_\theta\in \mathcal{G}} \E_{\eta \sim\pi_\theta}\bigg[\sum_{t=0}^{T} \gamma^t r_{t+1}\bigg],
\label{eq:rl}
\end{equation}
where $\mathcal{G}$ is the set of all candidates policies $\pi_\theta$, and $\eta$ is a sampled roll-out (i.e., a SMILES string) of length $T$ from $\pi_\theta$. $\gamma$ is the discount factor and usually $\gamma < 1$. Solving equation \eqref{eq:rl} is a difficult problem, and many algorithms have been proposed in past decades \cite{williams1992simple,schulman2017proximal,sutton2018reinforcement}. Here, we use a classical RL algorithm called REINFORCE.

\textbf{REINFORCE. }
REINFORCE belongs to a family of RL algorithms called policy gradient methods \cite{sutton2018reinforcement} which estimates the gradient of certain performance measures of a decision-making policy and inputs the gradient into a stochastic gradient ascent algorithm to improve the policy toward higher total rewards.
Formally, if $G_t=\sum_{k=t}^T \gamma^{k-t}r_{t+1}$, 
REINFORCE seeks to maximize  
$$\mathcal{J}(\theta) = \E\big[\log \pi_\theta(a_t|\bm{s}_t)G_t\big],$$
where $\mathcal{J}(\theta)$ captures the expected return under the distribution of all possible state and action sequences. Note that in REINFORCE, the policy is probabilistic, i.e., $\pi_\theta(a_t|s_t)$ is the probability of taking action $a_t$ in state $s_t$. Hence, the gradient of the objective function $\mathcal{J}(\theta)$ can be written as follows
$$\nabla\mathcal{J}(\theta)=\E\big[\nabla_\theta \log \pi_\theta(a_t|s_t)G_t\big].$$

However, computing $\nabla\mathcal{J}(\theta)$ is non-trivial due to the high dimensionality of the space of possible state and action sequences. REINFORCE addresses this problem by using Monte Carlo sampling and approximating the gradient by 
\begin{equation}
    \nabla\mathcal{J}(\theta)\approx \frac{1}{K}\sum_{i=1}^K \sum_{t=0}^{T_i}\nabla_\theta \log \pi_\theta(a_t^i|\bm{s}_t^i)G_t^i.
    \label{eq:grad_approx}
\end{equation}
At each iteration, REINFORCE samples $K$ roll-outs $\{\bm{s}_t^i, a_t^i, r^i_{t+1}\}_{i=1}^K$ from the current policy $\pi_\theta$ (i.e., the generative model), which are used to estimate $\nabla\mathcal{J}(\theta)$ using Equation \eqref{eq:grad_approx}. Then, parameters of the policy $\pi_\theta$ can be updated as
\begin{equation}
    \theta' = \theta + \alpha \nabla\mathcal{J}(\theta),
\end{equation}
where $\alpha$ is the learning rate.

Due to the high variance in the sampling process,  training can be unstable. To address this, a baseline reward $b_t^i$ is often estimated and subtracted from $G_t$. Hence, the gradient becomes
\begin{equation}
    \nabla\mathcal{J}(\theta)\approx \frac{1}{K}\sum_{i=1}^K \sum_{t=0}^{T_i}\nabla_\theta \log \pi_\theta(a_t^i|\bm{s}_t^i)(G_t^i-b^i_t)\big].
    \label{eq:reinforce_bs}
\end{equation}
Thus, REINFORCE can learn the parameters of the generative model in an end-to-end fashion by using backpropagation. During training, actions leading to higher total rewards $G_t$ will be reinforced through increasing $\log \pi_\theta$; while actions resulting in lower total reward will be suppressed by decreasing $\log \pi_\theta$.

\textbf{The Multi-Objective Reward Function. }
RL algorithms require a properly defined reward function. In this paper, we aim to learn a generative model that is able to generate SMILES strings with multiple desired properties: 1) smaller MW; 2) higher logP; 3) higher logS and 4) higher pIC50. Note that we only build predictors for logP, logS and pIC50, while MW is computed directly from the SMILES string. Therefore, we introduce a multi-objective reward function which is a weighted sum of these properties 
\begin{equation}
    r(\bm{s}_T)=\sum_{p\in \mathcal{P}} w_p r_p(\bm{s}_T),
    \label{eq:weighted_rwd}
\end{equation}
where $\mathcal{P}=\{\text{MW, logP, logS, pIC50}\}$, $w_p$ is the weight assigned to each property and $r(\bm{s}_T)$ is the predicted property value for the generated SMILES string $\bm{s}_T$. Besides, we assign a negative weight to MW to convert the minimization task to maximization.
In addition, to ensure the validity of most generated SMILES strings, we regularize the generative model by penalizing the model when it  generates an invalid SMILES string, with a negative reward $r_p$. Hence, we define the reward function $R(\bm{s}_t, a)$ as follows:
\begin{align}
    R(\bm{s}_t, a) = \left\{ \begin{array}{ll} 
                0 & \text{if } t<T \\
                r(\bm{s}_T) & \text{if } t=T \text{ and } \bm{s}_t\text{ is valid} \\
                r_p & \text{otherwise}\\
                \end{array} \right.
    \label{eq:rwd_fun}
\end{align}
The reward is only provided at the $T$th (last) step of the generation. REINFORCE uses this reward function to learn a customized generative model that is capable of generating valid SMILES strings with multiple desired properties.

\subsection{Data Collection }
In order to train the generative model, we set up a SMILES-strings corpus with 1,870,310 unique compounds, which is retrieved from ChEMBL25 database\cite{gaulton2017chembl}. Note that all SMILES strings here are canonicalized, which means that they are uniquely mapped to compounds. 
To train the predictive model, we acquire logS and logP data from the literature\cite{sorkun2019aqsoldb, popova2018deep} and removed the duplicates. IC50 data against MOR (ChEMBL ID: 4354) are retrieved from ChEMBL25 database\cite{gaulton2017chembl}. We only include compounds with explicit IC50 values at the same scale. If the IC50 value is low, then the corresponding compound is highly potent against its target since only a very little amount of the compound can inhibit the target. We take their negative logarithm to get the pIC50 dataset. A high pIC50 means that the compound has high potency.
Basic statistics for the three datasets are summarized in Table~\ref{number of unique compounds}.

\begin{table}[b]
\centering
  \caption{Statistics for the Datasets in the Predictive Model}
  \label{number of unique compounds}
  \begin{tabular}{l|llll}
    \hline
    Property  & Min & Max & Median & Count \\
    \hline
    logP & -5.1 & 11.3 & 2.0 & 14,152  \\
    logS & -13.2 & 2.1 & -2.6 & 9,981  \\
    pIC50 & 1.8 & 10.2 & 6.1 &915  \\
    \hline
  \end{tabular}
\end{table}

\subsection{Model Architecture}
\begin{figure}[h!]
\centering
\includegraphics[width=0.99\linewidth]{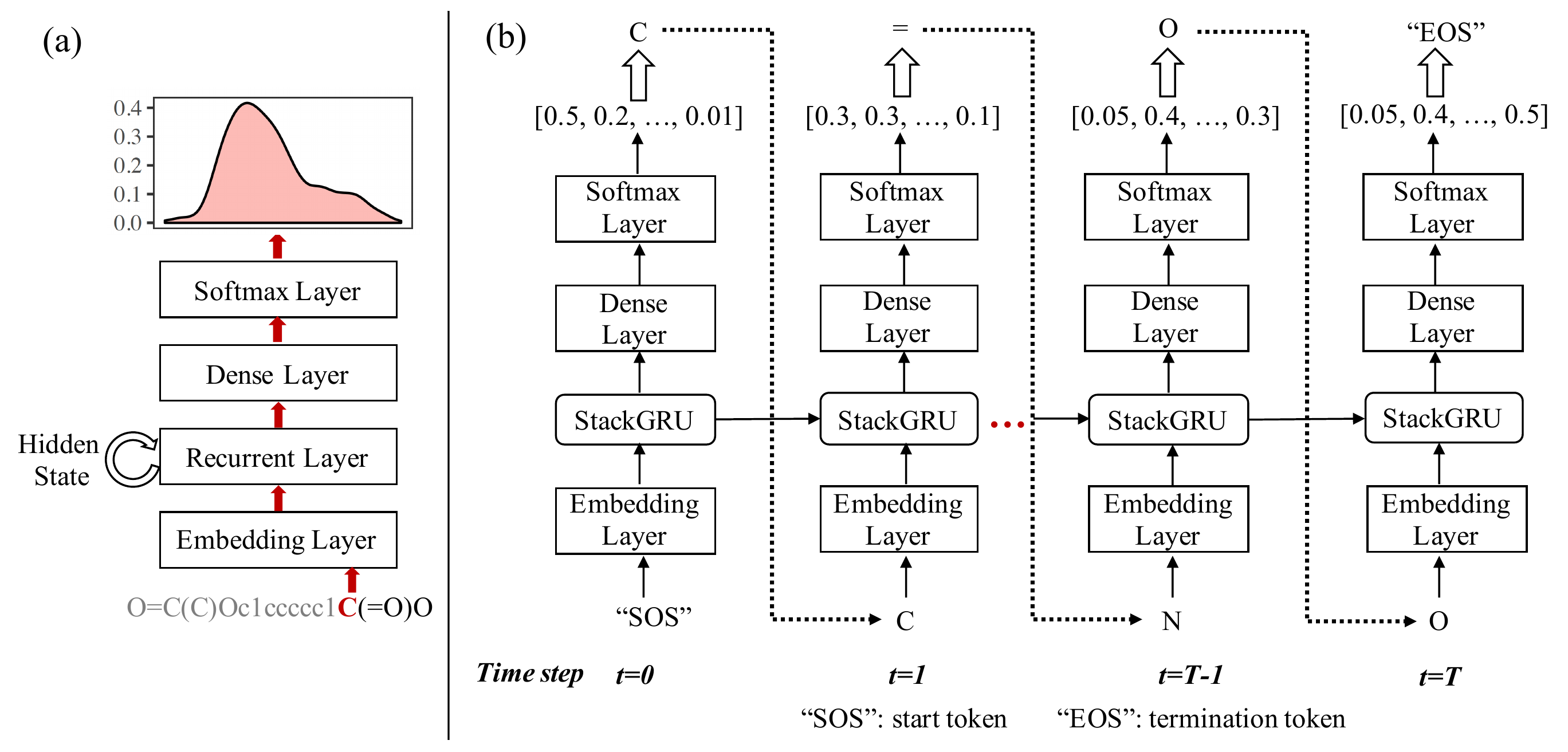}
\vspace{-.1in}
\caption{The Generative Model. (a) Model architecture. (b) Illustration of the generation of a SMILES string.}
\label{fig:Generative_Model}
\end{figure}

\textbf{The Generative Model. }
To accelerate the training of the generative model in the RL stage, we first train the generative model to learn the syntactical rules for constructing SMILES strings.
At each time step, the generative model takes a current prefix string of a training instance (i.e., a SMILES string), and predicts the probability distribution of the next character (Figure~\ref{fig:Generative_Model}(a)). A cross-entropy loss is calculated at each step and parameters of the model are updated through back propagation. By treating each step as a multi-class classification problem, we fit the generative model to existing SMILES strings such that the model can generate valid SMILES strings. Importantly, the learned weights later serve as a good initialization for the generative model and expedite the training in the RL stage.

The generative model is a recurrent neural network \cite{lecun2015deep} which is capable of generating SMILES strings. Essentially, a SMILES string is composed of a sequence of characters $\bm{s}=\{c_1, \cdots, c_T\}$ from a vocabulary $V$ and $c_t\in V$ for all $t\in [1,T]$. To facilitate the generation process, we append a start token and a termination token to the head and tail of $\bm{s}$, respectively. As shown in Figure~\ref{fig:Generative_Model}(b), at each time step $t\in[0, T]$, the input to the generative model is a character $c_t$ ($c_0$ is the start token). The generative model first uses an embedding layer to convert the categorical character $c_t$ into an embedding vector of continuous scalars, which is then processed by a recurrent layer to update its hidden state. Finally, a dense layer and a softmax layer are used to map the hidden state to a probability distribution of the next possible character, from which we sample the next character. By repeating this process until a termination token is sampled, the generative model generates a complete SMILES string.

\begin{figure}[t]
\centering
\includegraphics[scale=0.5]{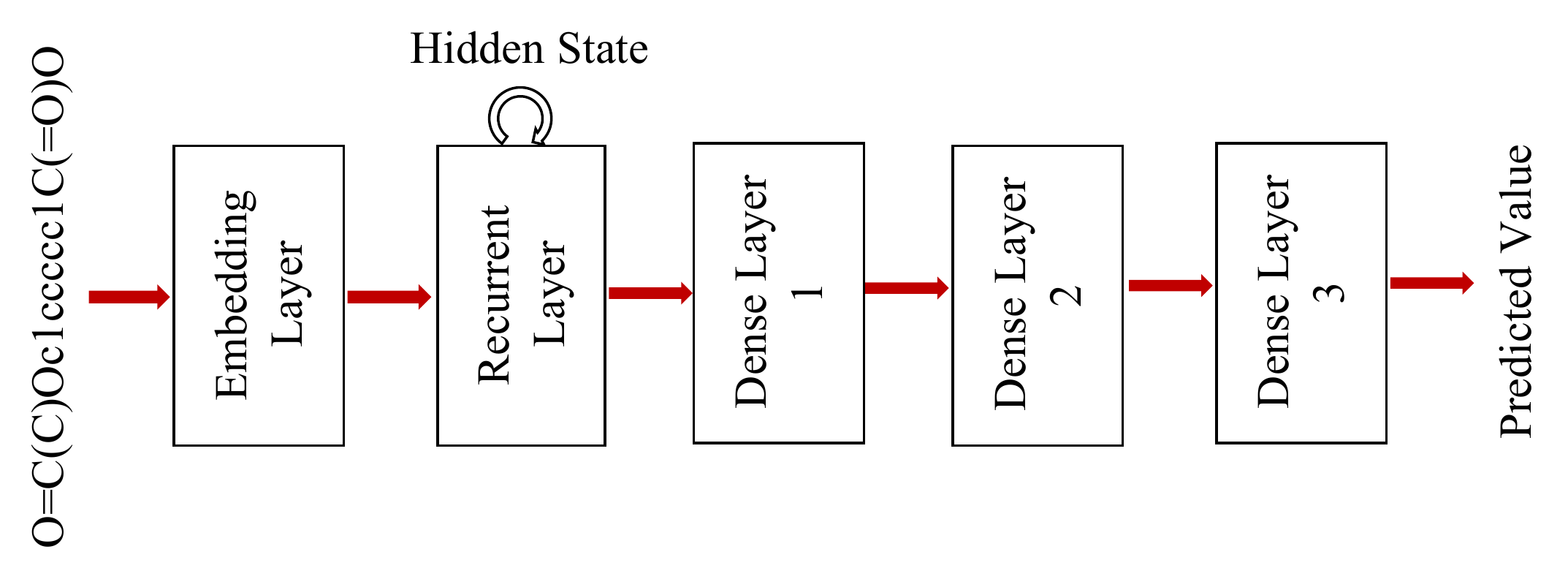}
\vspace{-.2in}
\caption{The Architecture of the Predictive Model.}
\label{Predictive_Model_DL}
\end{figure}

\textbf{The Predictive Model. }
The predictive model consists of multiple property predictors, each for one property of interest. Here, we consider three properties, namely, logP, logS and pIC50. Property prediction is essentially a regression task where we aim to map a SMILES string to a scalar value. All property predictors are RNN-based models with the same architecture. Figure \ref{Predictive_Model_DL} illustrates the  predictive model architecture. First, a SMILES string is passes through an embedding layer, converting each character in the SMILES string into an embedding vector. Second, the recurrent layer sequentially processes the embedding vectors and constructs a temporal feature vector for the input SMILES string. Last, three consecutive dense layers are used to map the feature vector to a property value. The network is trained with a mean squared error (MSE) loss.

However, a large number of training examples are often required for deep learning models, like RNN, before they can achieve superior performance. For cases where only limited training examples are available, Support Vector Machines \cite{cortes1995support} and Random Forests \cite{liaw2002classification} are often more competitive. Hence,
for each property, we compare three different models: Support Vector Machines \cite{cortes1995support}, Random Forests \cite{liaw2002classification} and the proposed RNN-based model, and select the classifier with the smallest MSE. 
We find Random Forest works best for pIC50 prediction; while the RNN-based model works best for logS and logP prediction. 
This can be due to the fact that we only have a small number of data points ($\sim$1k) for pIC50. In contrast, $\sim$14k and $\sim$10k training examples are available for logP and logS, respectively.

\subsection{Implementation Details.}
In Equation \eqref{eq:rwd_fun}, we set the penalty of invalid SMILES to $r_p=\hat{r}_\mu - \hat{r}_\sigma$, where $\hat{r}_\mu$ and $\hat{r}_\sigma$ are the average and standard deviation of the weighted sum of rewards (i.e., $r(\bm{s}_T)$ in Equation \eqref{eq:weighted_rwd}) of the SMILES strings sampled from the initialized generative model (prior to the RL training stage). Importantly, we empirically find that the RL algorithm is sensitive to the $r_p$ and our setting of $r_p$ strikes a good balance between validity and desired properties. We assign an equal weight of $0.25$ to logP, logS and pIC50, and $-0.25$ to MW.
During the RL training stage, the learning rate and maximum length of SMILES strings are set to $10^{-5}$ and $200$, respectively. The generative model is trained for 240 episodes. In each episode, we sample 200 SMILES strings with a batch size of 10.

The size of the dictionary is 58 with 56 distinct SMILES tokens plus a start token and a termination token. Each token is embedded into a 512-dimensional vector. 
Traditional RNN models like GRU \cite{cho2014gru} and LSTM \cite{Hochreiter-Schmidhuber-NC97} are inferior in memorizing and counting thus often fail to capture algorithmic patterns in sequences \cite{joulin2015inferring}. Meanwhile, the construction of SMILES strings needs to follow certain algorithmic rules such as atom valence constraints and bracket opening-closure. Thus, here we use Stack-augmented GRU (StackGRU) \cite{joulin2015inferring} as the recurrent layer in the generative model. The stack width, stack depth and hidden size of StackRNN are 256, 200, 512, respectively.
For the predictive model, an ordinary single-layer GRU with hidden size 512 is used as the recurrent layer. The following three dense layers have dimension sizes 128, 32, and 1 respectively, the first two of which are followed by a ReLU layer and a batch-norm layer \cite{ioffe2015batch}. The deep learning models are implemented with PyTorch \cite{paszke2019pytorch}. RDKit \cite{landrum2006rdkit} is used for validating and visualizing the molecules. The Random Forest model for predicting pIC50 has $100$ trees and is implemented with Scikit-Learn \cite{pedregosa2011scikit}. 

\section{Results}

\subsection{Evaluation of the DRL Framework}
\textbf{GRU vs StackGRU.}
We first compare the GRU and StackGRU in learning the syntax of SMILES strings. Specifically, we train two generative models (see Figure \ref{fig:Generative_Model}(a): one with GRU as the recurrent layer; the other with StackGRU) on the training corpus, and sample 10k SMILES strings from each model.
By using syntactical and chemical validity check functions from RDKit, we calculate the percentage of syntactically and chemically valid compounds. 
We also measure the novelty of generated compounds by calculating the percentage of non-overlapping compounds between the sample and the training corpus. Furthermore, by examining the percentage of non-duplicates within the sample, we quantify the uniqueness of the generated sample.

\begin{table}[b]
\centering
  \caption{Performance Comparison between GRU and StackGRU}
  \label{performance of generative model}
  \begin{tabular}{l|llll}
    \hline
    Configuration & Syntactical Validity (\%) & Chemical Validity (\%) & Novelty (\%) & Uniqueness (\%) \\
    \hline
    GRU & 75.87 & 60.74 & 99.32 &99.99   \\
    StackGRU & 87.29 & 77.40 & 98.92 & 99.97  \\
    \hline
  \end{tabular}
\end{table}

As can be seen from Table~\ref{performance of generative model}, both syntactical validity and chemical validity are relatively low when using a standard GRU compared to the StackGRU.  With StackGRU, syntactical validity increased to 87.29\% and chemical validity increased to 77.40\%, which demonstrates that StackGRU is better at learning the SMILES syntax. Both novelty and uniqueness are close to 100\%, which indicates that the generative model is able to generate novel and unique SMILES strings, and does not just memorize training examples. 

\begin{figure}[h]
\centering
\includegraphics[scale=0.55]{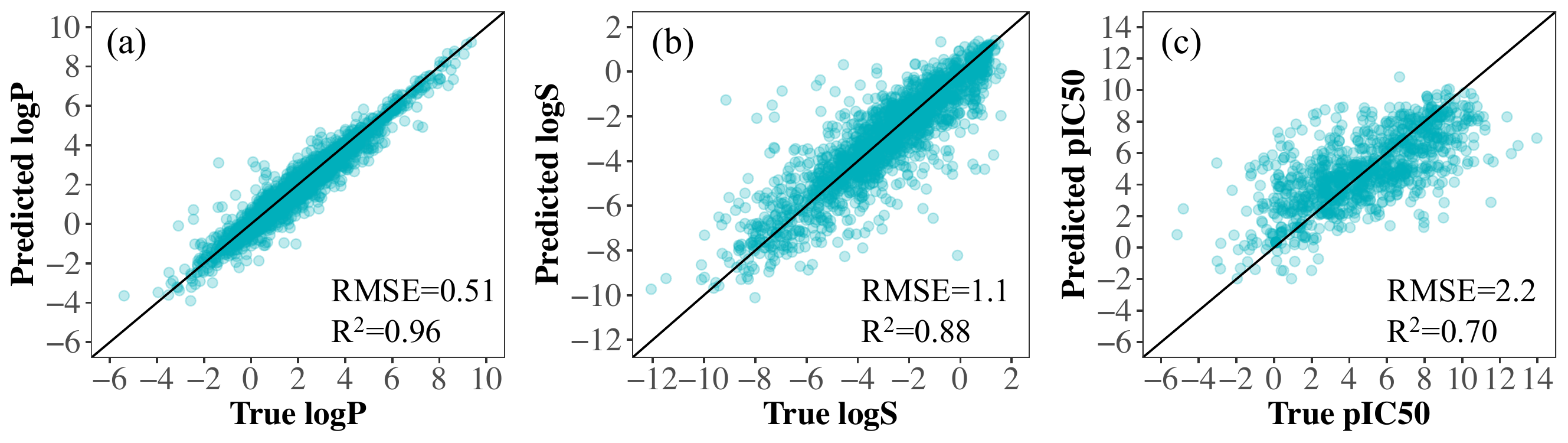}
\caption{Predicted Value vs True Value from the Predictive Model.}
\label{Prediction_Performance}
\end{figure}

\textbf{Property Prediction.}
For the predictive model, we plot the predicted value vs true value in Figure ~\ref{Prediction_Performance}. 
Indicated by a high correlation coefficient, i.e., $R^2$ and low rooted mean squared error (RMSE), the predictors for logP and logS have high precision and accuracy. However, the sub-predictive model for pIC50 has relatively poor performance, which can be caused by the insufficient data points in the pIC50 dataset. We also apply the predictive model to Naloxone. The predicted values for logP and logS of Naloxone are 1.94 and -2.67, respectively, which align reasonably well with the reported properties of Naloxone (logP=1.47 and logS=-1.8) \cite{wishart2018drugbank}.

\textbf{Molecule Generation. }
Given that our goal is to bias the generation of molecules towards higher logP, logS, pIC50 and smaller MW, we sample 10k SMILES strings after the generative model is trained using REINFORCE for 0, 80, 160 and 240 episodes, respectively.

\begin{figure}[h]
\centering
\includegraphics[width=0.9\linewidth]{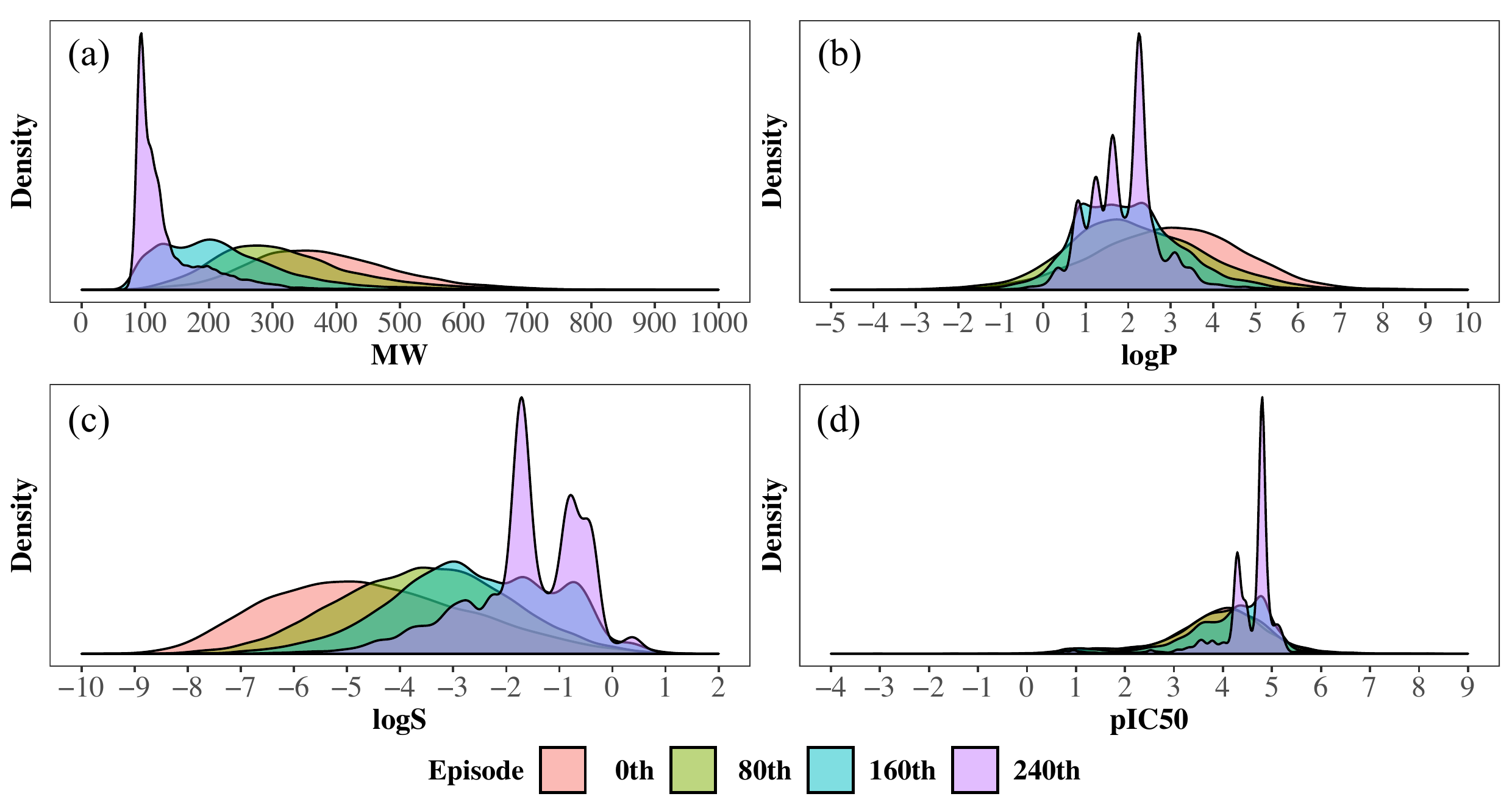}
\caption{Shifted Distribution of Targeted Properties during RL Training. }
\label{performance_4multi_RL}
\end{figure}

Figure~\ref{performance_4multi_RL} shows the distribution of each property value for chemically valid strings from the samples. The  pIC50 of the  generated molecules are biased toward higher values, albeit not significantly. The distribution of logS is significantly shifted to the right and the distribution of MW is also significantly shifted to the left, which means molecules with higher logS and smaller MW are more likely to be generated over training episodes. However,  generated molecules tend to have lower logP values, indicated by the left-shifted distribution of logP. This phenomenon is probably because logP and logS are contradictory by nature. When there are more hydrophilic groups, higher logS and lower logP are expected and \textit{vice versa} when there are more hydrophobic groups. Overall, our DRL framework is able to bias the properties of generated molecules.

Table~\ref{performance of multiRL model} summarizes the syntactical validity, chemical validity, novelty and uniqueness of the generated samples. 
\begin{table}[b]
\centering
  \caption{Evaluation of Generated Samples during RL Training}
  \label{performance of multiRL model}
  \begin{tabular}{l|llll}
    \hline
    Episode & Syntactical Validity (\%) & Chemical Validity (\%) & Novelty (\%) & Uniqueness (\%) \\
    \hline
    0th & 87.29 & 77.40 & 98.92 & 99.97   \\
    80th & 91.92 & 89.60 & 95.58 & 98.50  \\
    160th & 96.77 & 95.90 & 93.79 & 65.52  \\
    240th & 99.40 & 99.16 & 96.46 & 21.51  \\
    \hline
  \end{tabular}
\end{table}
One major issue with RL in \textit{de novo} drug design in previous studies is the reduced validity\cite{popova2018deep}. Here, by incorporating penalty for invalid SMILES strings in the reward function, our DRL framework generates SMILES strings approaching 100\% validity when the training episodes increase. Besides, the novelty of the generated samples is also high. Uniqueness is decreasing as the number of episodes goes up since as the DRL training episodes increase, the generative model tends to converge to the distribution of a smaller number of SMILES strings with the desired properties.

\subsection{Identification of Potential Lead Compounds}
From the RL-trained generative model at the 160th episode, we sample 10k SMILES strings and then use the following criteria to filter out molecules with: 1) logP $>1.94$, 2) logS $> -2.67$, 3) pIC50 $>6$ and 4) MW $< 327.37$. The reason why we choose the 160th episode is that it enables balanced performance with regard to the validity, novelty, uniqueness and the ability to shift property distribution. The cutoff values used in logP, logS and MW  are the predicted/calculated values for naloxone in our DRL framework. Note that our goal is to discover molecules with sufficient inhibitory activity against MOR and optimal logP, logS and MW values to ensure prolonged brain retention ability. We set the cutoff value of pIC50 at 6 despite the predicted pIC50 for naloxone by our predictive model is 6.93 since pIC50 $\geq 6$ corresponds to active compounds\cite{popova2018deep}. 
\begin{figure}[h]
\centering
\includegraphics[width=0.75\linewidth]{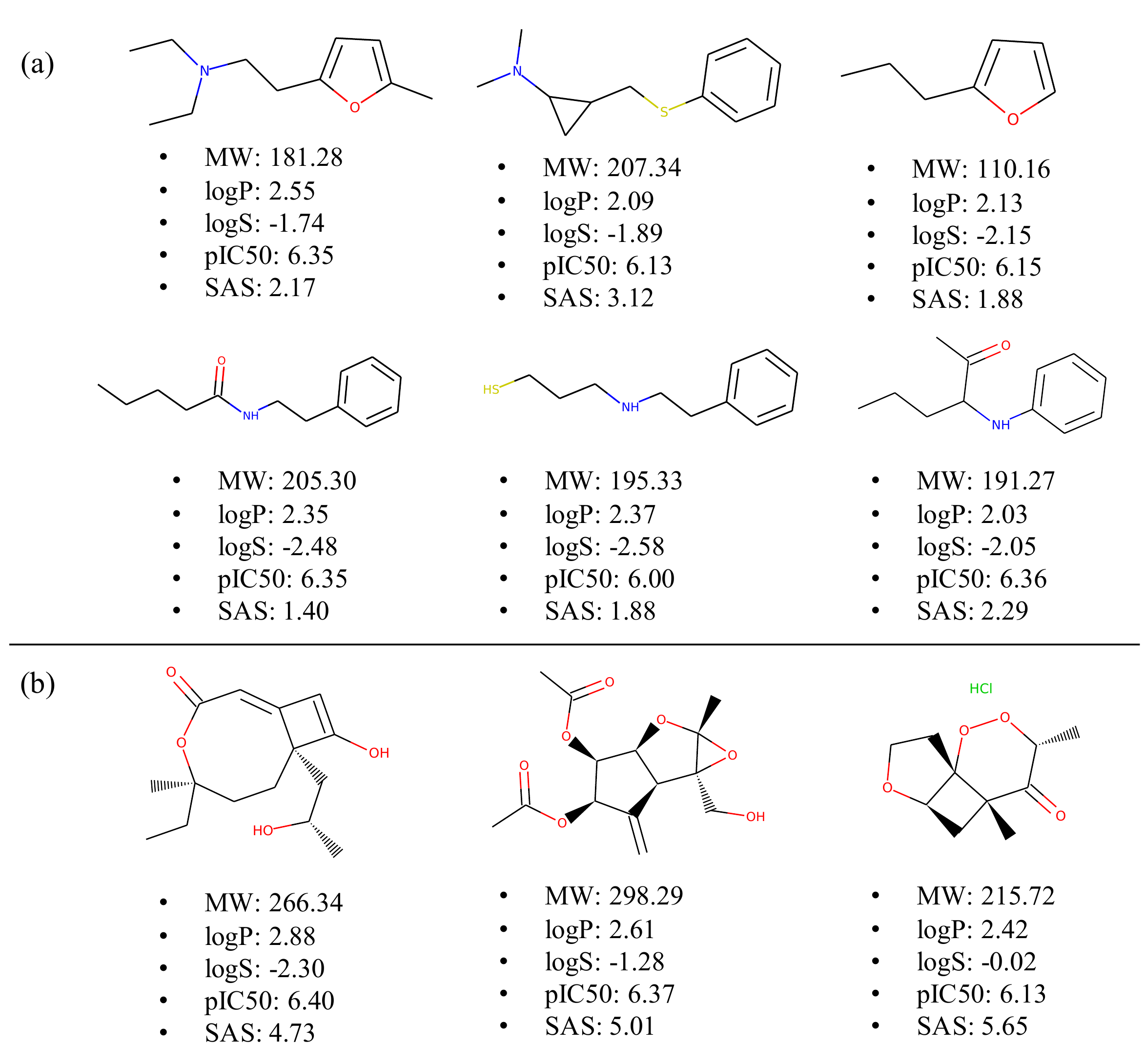}
\caption{Novel Molecules Generated in the DRL Framework. (a) At the 160th Episode. (b) At the 0th Episode.}
\label{generated_sructures}
\end{figure} 

We filtered out six novel SMILES strings  from the 10k-size sample, as the identified potential lead compounds. Figure~\ref{generated_sructures} (a) shows their structures drawn by RDKit, which are simple by direct visual checking. We also calculated their synthetic accessibility score (SAS)\cite{ertl2009estimation}. SAS can range between 1 and 10. A high SAS, usually above 6, corresponds to high molecule complexity and increased synthesis difficulty. For the six molecules, their SAS values range from 1.40 to 3.12, indicating that our DRL framework generates highly feasible molecules. 

To further demonstrate the usefulness of the DRL framework, we also sample 10k SMILES strings at the 0th episode (i.e., without RL training) and filter with  the same criteria. Three SMILES strings are filtered out. Figure~\ref{generated_sructures} (b) shows their structures and predicted properties. Despite having the expected properties, their structures are very complex with high SAS values, which indicates that the molecules generated without RL training are much less feasible. 

\section{Conclusion and Discussion} 
Recent years have seen an increasing use of deep learning in drug discovery with the rise of the `big data' era\cite{chen2018rise}. Linear representations of molecules, such as the SMILES strings, are broadly used in ligand-based drug discovery studies\cite{lipinski2019advances}. By linking molecules to end points like physicochemical properties, inhibitory activity, BBB permeability, etc, end-to-end drug discovery and development is now becoming a reality  \cite{ekins2019exploiting}.
For example, in 2019, Insilico Medicine succeeded in using deep learning to design new lead compounds for discoidin domain receptor 1 (DDR1) kinase inhibitors from scratch in just 21 days\cite{zhavoronkov2019deep}. 

In this study, we aim to discover better opioid antagonists to help to combat the opioid epidemic, where the most common antidote for opioid overdose, naloxone, has limited blood brain barrier permeability.
To accelerate the discovery for better opioid antogonists candidates, we implement a multi-objective DRL framework. The framework is able to identify valid, novel and feasible molecules with sufficient opioid antagonistic activity and enhanced brain retention ability. 

More importantly, the proposed multi-objective DRL framework has great potential in accelerating drug discovery, which is a multi-property optimization task \textit{per se}. For instance,  effective and safe drugs need to exhibit a fine-tuned combination of pharmacokinetic and pharmacodynamic properties, such as high potency, affinity and selectivity against the drug target as well as optimal absorption, distribution, metabolism, excretion and toxicity (ADMET)\cite{ferreira2019admet}. Our study shows that with well-designed reward functions, the multi-objective DRL framework can be customized to generate molecules with optimal properties from different drug development aspects.

\bibliographystyle{unsrt}
\bibliography{reference}

\end{document}